\documentclass[aps,prl,twocolumn,superscriptaddress,groupedaddress,english]{revtex4}
\usepackage[T1]{fontenc}
\usepackage[utf8]{inputenc}
\usepackage{amsmath}
\usepackage{graphicx}
\usepackage{setspace}
\usepackage{babel}
\usepackage{amssymb}
\usepackage{leftidx}
\usepackage[titletoc,toc,title]{appendix}
\usepackage{titlesec}
\usepackage{braket}
\usepackage{bm}
\usepackage{subfig}
\usepackage{xcolor}
\colorlet{yy}{green!10!orange!100!}
\colorlet{yr}{green!10!blue!80!}
\colorlet{yg}{green!65!blue!90!}
\colorlet{yp}{purple!80!blue!80!}
\def\iu{\mathrm{i}}
\usepackage{caption}
\colorlet{shadecolor2}{yellow!10}
\colorlet{shadecolor}{gray!10}
\usepackage[pdftitle={Draft paper},pdfauthor={Giammarco Fabiani}]{hyperref}
\hypersetup{colorlinks=flase}
\usepackage{titlesec}
\titleformat{\section}[runin]
  {\normalfont}{\thesection}{1em}{}
\titleformat{\subsection}[runin]
  {\normalfont\bfseries}{\thesubsection}{1em}{} 
\begin{document}
\title{Supermagnonic propagation in two-dimensional antiferromagnets}
\author{G. Fabiani} \author{M. D. Bouman} \author{J. H. Mentink}
\affiliation{Radboud University, Institute for Molecules and Materials (IMM) Heyendaalseweg 135, 6525 AJ Nijmegen, The Netherlands}
\date{\today}

\begin{abstract}
We investigate the propagation of magnons after ultrashort perturbations of the exchange interaction in the prototype two-dimensional Heisenberg antiferromagnet. Using the recently proposed neural quantum states, we predict highly anisotropic spreading in space constrained by the symmetry of the perturbation. Interestingly, the propagation speed at the shortest length and time scale is up to 40\% higher than the highest magnon velocity. We argue that the enhancement stems from extraordinary strong magnon-magnon interactions, suggesting new avenues for manipulating information transfer on ultrashort length and time scales.
\end{abstract}
\maketitle

\section{\emph{Introduction.}}

The study of magnons, the collective spin excitations in magnetic systems, has triggered significant interest in recent years. Stimulated by the potential for high-speed low-energy data processing, high-energy coherent magnons are intensively investigated in the field of magnonics \cite{Chumak 2015} and spintronics \cite{Grundler, Wang 2020}. In addition, high-energy magnons have a crucial role in the microscopic dynamics of ultrafast switching between magnetically ordered states \cite{Radu, Graves, Iacocca, Ruta, Buttner}, and are potentially essential for the stabilization of various complex quantum many-body states \cite{Novelli, Dean 2016, Mazzone}. Furthermore, these studies are greatly stimulated by the availability of femtosecond X-ray techniques \cite{Buzzi}, which ultimately can measure the propagation of magnons with nanometer spatial and femtosecond temporal resolution. Nevertheless, rather little is known about the propagation of high-energy magnons at these ultrashort length and time scales.
 
A direct way to access high-energy magnons is via optical perturbations of the exchange interactions \cite{Mentink 2017}, as well established in Raman spectroscopy, both in the frequency \cite{Weber, Deveraux} and time domain \cite{Zhao, Bossini 2016, Bossini 2019}. In this approach, high-energy magnons are excited in pairs with wavelengths as small as the distance between two atoms, corresponding to oscillation frequencies determined by the exchange energy. 
Interestingly, the spectrum of these magnon pairs is significantly affected by magnon-magnon interactions \cite{Elliot, Canali, Lorenzana}. This is particularly true for the case with strongest quantum spin fluctuations, i.e., the relevant case of spin $S=1/2$ in two dimensions (2D) \cite{Canali, Sandvik 1998}, for which even the single-magnon spectra are strongly modified at short wavelengths \cite{Coldea, Christensen, Headings, Dean 2012, Le Tacon, Powalski 2015, Powalski 2018, Dalla Piazza, Sandvik 2017}. Hence, magnon-magnon interactions might have a pronounced effect on the propagation of high-energy magnons, especially in the systems for which experimentally the strongest quantum fluctuations are found \cite{Lyons, Singh}.
Therefore, we aim to understand both \textit{how} magnon-magnon interactions influence the propagation of magnon pairs, and to quantify \textit{how strong} this effect becomes in the regime of strongest quantum fluctuations.

Theoretical investigation of magnon propagation in this deep quantum regime is highly challenging, since it requires to solve the unitary dynamics of an extended quantum many-body system with strong spatial and temporal quantum spin correlations, for which no exact methods exist. Recently, however, a new family of algorithms was proposed which are inspired by machine learning \cite{Carleo}. Although being inherently a variational method, these neural quantum states (NQS) offer a nearly unbiased approach to the full quantum dynamics. In particular, it was shown that NQS are highly efficient for the simulation of quantum spin dynamics in the most challenging 2D limit \cite{Fabiani, Schmitt}. 

Here, we apply the NQS to investigate the propagation of magnons after ultrashort perturbations of the exchange interaction in the square lattice spin-1/2 antiferromagnetic Heisenberg model. We find that the correlation spreading resembles the anisotropic propagation pattern expected from non-interacting magnons. Interestingly, however, at the shortest length and time scales, we predict that the spreading speed qualitatively deviates from non-interacting magnons, reaching speeds that are significantly higher than the highest magnon group or phase velocity. By comparison with approximate results obtained with Schwinger boson mean-field theory, we identify that this enhanced spreading speed stems from an interplay between a localized quasi-bound state emerging from magnon-magnon interactions and propagating, nearly non-interacting magnon pairs. We predict 40\% enhancement of the propagation speed in the regime of strongest quantum fluctuations.

\section{\emph{Model and method.}}

We study the spin-$1/2$ antiferromagnetic Heisenberg model on a square lattice with $N=L\times L$ spins $\hat{\mathbf{S}}_i=\hat{\mathbf{S}}(\mathbf{r}_i)$, with $\mathbf{r}_i=(x_i,y_i)$ 
\begin{equation}\label{eq:1}
\hat{\mathcal{H}} = J_{\text{ex}} \sum_{\langle ij\rangle } \hat{\mathbf{S}}_i \cdot \hat{\mathbf{S}}_{j},
\end{equation}
where $J_{\text{ex}}$ is the exchange interaction ($J_{\text{ex}}>0$) and $\langle \cdot \rangle$ restricts the sum to nearest neighbours. 
We consider the dynamics induced by a time-dependent perturbation of the exchange interaction \cite{Zhao, Bossini 2016, Bossini 2019, Mentink 2015, Mentink 2017, Rostislav}, modeled by the perturbation
\begin{equation}\label{eq:2}
\delta \hat{\mathcal{H}}(t) = \Delta J_{\text{ex}}(t)\frac{1}{2} \sum_{i,\bm{\delta}}\big(\mathbf{e}\cdot \bm{\delta}\,\big)^2 \,\hat{\mathbf{S}}(\mathbf{r}_i) \cdot \hat{\mathbf{S}}(\mathbf{r}_i+\bm{\delta}),
\end{equation}
where $\mathbf{e}$ is a unit vector that determines the polarization of the electric field of the light pulse which causes the perturbation and $\bm{\delta}$ connects nearest neighbour spins. This perturbation is reminiscent to the Loudon-Fleury theory of spontaneous Raman scattering \cite{Fleury}. In the remainder of this work we set $\hbar=1$ and the lattice constant $a=1$, and work at zero temperature.

To simulate the real-time dynamics following Eq. \eqref{eq:2} we employ the recently introduced neural quantum state ansatz inspired by machine learning \cite{Carleo}. This approximates the wavefunction of the system with a restricted Boltzmann machine (RBM) which can be expressed as 
\begin{equation}\label{eq:RBM}
\psi = \text{exp}\Big(\sum_i a_i S^z_i\Big) \prod^M_{i=1} 2\,\text{cosh}\Big(b_i+\sum_j W_{ij}S^z_j\Big).
\end{equation}
Here $S_i^z=\pm \,1/2$ correspond to the physical spins and $\{a_i,b_i,W_{ij}\}$ are complex coefficients that parametrize the many-body wavefunction. The number of variational parameters is $N_{\text{var}} = \alpha \times N^2 + \alpha \times N + N$, with $\alpha=M/N$ controlling the accuracy of the ansatz. The neural network is trained by means of variational Monte Carlo techniques to simulate the ground state and time-dependent states of a given lattice Hamiltonian. In particular, unitary dynamics is addressed by employing the time-dependent variational principle  \cite{Carleo 2012}.

In a previous work we showed that the RBM ansatz can reproduce the ground-state properties of Eq. \eqref{eq:1} and the dynamic properties under Eq. \eqref{eq:2} with high accuracy \cite{Fabiani}. Here we adopt a similar protocol approximating the time-dependent change of the exchange interaction as a global quench of $J_{\text{ex}}$ along ${\mathbf{e}}=\mathbf{y}$ with $\Delta J_{\text{ex}}(t)= 0.1\, J_{\text{ex}}\,\Theta(t)$, where $\Theta(t)$ is the Heaviside function. For the short-time dynamics considered here, this closely resembles the square pulse protocol adopted in \cite{Fabiani}.
Our numerical simulations always start from the ground state of Eq. \eqref{eq:1} and are obtained using the ULTRAFAST code \cite{Fabiani}.

\section{\emph{Results.}}

\begin{figure}  
 \vspace*{-14pt}\hspace*{-20pt} \subfloat{{\includegraphics[width=8.6cm]{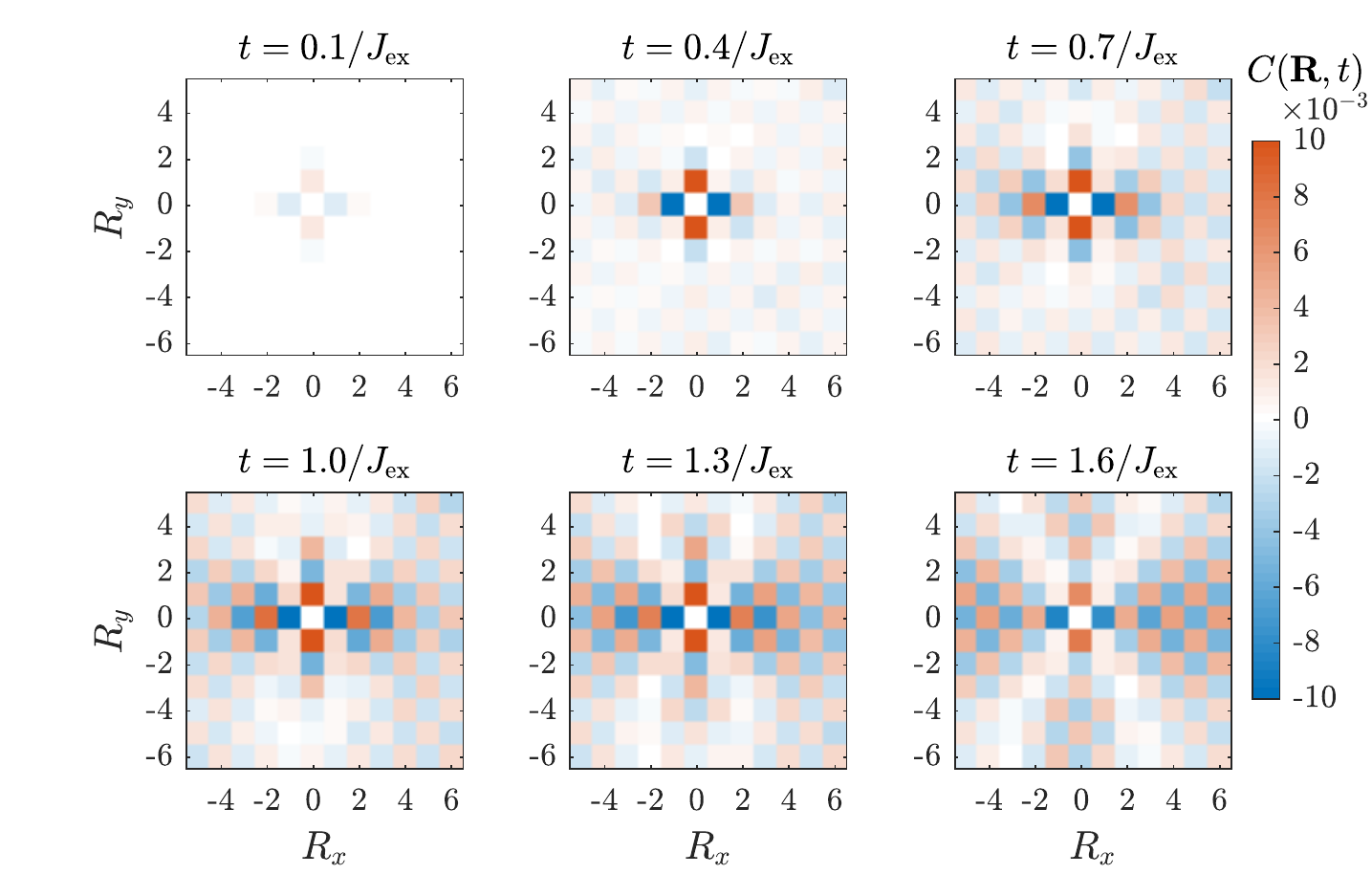}}}
    \caption{(Color online) Snapshots in time of the post-quench dynamics of spin correlations in an $L\times L=12\times 12$ system simulated with the RBM ansatz ($\alpha=16$). The figures reveal a clear spreading of correlations, with a non-trivial wavefront developing at ultrashort time scales. The checkerboard pattern reflects the antiferromagnetic coupling between spins. The noise in the correlations is due to Monte Carlo errors, and to improve readability the colormap is clipped at $\pm 10^{-2}$.}
    \label{fig:spreading}
\end{figure} 

According to a well established quasiparticle picture \cite{Calabrese}, the space-time dynamics of quantum correlations after a quench is determined by the quasiparticles excited. Therefore, to investigate the propagation of magnons triggered by Eq. \eqref{eq:2}, we consider the time evolution of 
\begin{align}\label{eq:corr}
C(\mathbf{R},t) = \big\langle\hat{\mathbf{S}}_i(t) \cdot \hat{\mathbf{S}}_j(t)\big\rangle_{\text{c}} - \big\langle\hat{\mathbf{S}}_i(0) \cdot \hat{\mathbf{S}}_j(0)\big\rangle_{\text{c}},
\end{align}
where $\langle\hat{A} \hat{B}\rangle_{\text{c}} = \langle\hat{A} \hat{B}\rangle - \langle\hat{A}\rangle \langle\hat{B}\rangle$
and $\mathbf{R}=\mathbf{r}_i-\mathbf{r}_j$. Both the system and the perturbation are translationally invariant and therefore the correlation function only depends on the relative distance $\mathbf{R}$ between the sites considered. 
Fig. \ref{fig:spreading} shows different snapshots in time of the correlator $C(\mathbf{R},t)$ obtained with the RBM ansatz in a $12\times 12$ lattice. We note that after $t\approx 1.6/J_{\text{ex}}$ the wavefront reaches the lattice boundaries and the subsequent spreading, dominated by finite-size effects, is not considered. Fig. \ref{fig:spreading} reveals a propagation pattern arising at very small time scales with a highly anisotropic wavefront, with almost vanishing correlations along the diagonals. The weak spreading along the diagonals derives from an exact symmetry of $C(\mathbf{R},t)$ that holds in the linear response limit $\Delta J_{\text{ex}}\ll J_{\text{ex}}$ (see Supplemental Material I). In this limit, the correlation function $C(\mathbf{R},t)$ is antisymmetric with respect to reflections over one of the diagonals of the lattice. As a consequence, $C(\mathbf{R},t)$ vanishes when $R_x= \pm R_y$. Small corrections beyond linear response break this symmetry, yielding a slight anisotropy between $\mathbf{x}$- and $\mathbf{y}$-axes, with finite (but small) correlations along the diagonals consistent with the spreading patterns of Fig. \ref{fig:spreading}.

\begin{figure}
    \vspace*{10pt}\centering
    \subfloat{{\includegraphics[width=8.6cm]{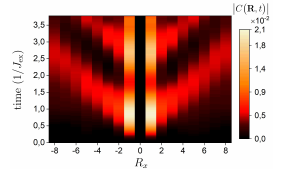}}}
    \caption{(Color online) Time evolution of spin correlations $|C(\mathbf{R},t)|$ as a function of the distance $R_x$ in an $L \times L=20\times 20$ system. A light-cone like spreading of correlations appears in agreement with the locality of $\hat{\mathcal{H}}+\delta \hat{\mathcal{H}}$. Results are obtained with the RBM ansatz using $\alpha=12$. }
    \label{fig:spreading1d}
\end{figure}

\begin{figure}
    \vspace*{10pt}\centering
    \subfloat{{\includegraphics[width=8.6cm]{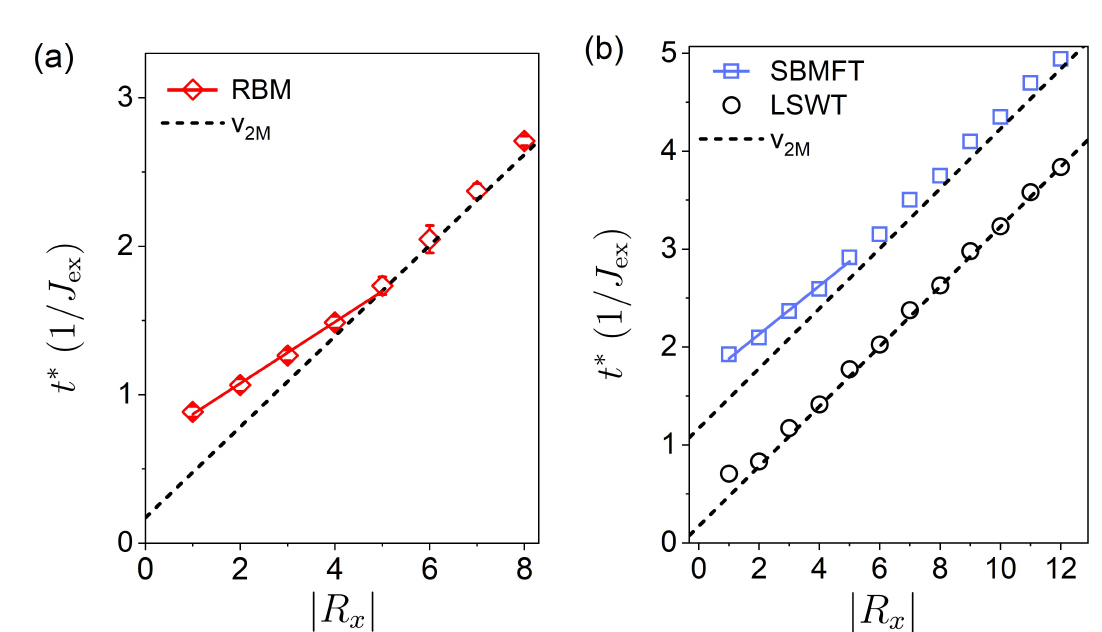}}}
    \caption{(Color online) Arrival times $t^*$ of the first extrema of $C(\mathbf{R},t)$. (a) RBM data (red diamonds) compared with the highest magnon group velocity $v_{\text{2M}}$ (black dashed line). (b) Comparison between SBMFT (blue squares) and LSWT (black circles). Solid red (blue) line: fit of the $|R_x|\leq 5$ arrival times of the RBM (SBMFT) correlations. To improve readability, the SBMFT data have been shifted in time  by $+1/J_{\text{ex}}$. The RBM points are calculated for $L\times L=20 \times 20$ ($\alpha=12$), while the LSWT and SBMFT data refer to an $L\times L=140\times 140$ system, where finite-size convergence is found.}
    \label{fig:maxima}
\end{figure}

In order to extract the speed of the correlation spreading, we focus on the correlations along the $\mathbf{x}$-direction with $\mathbf{R}=(R_x,0)$. Fig. \ref{fig:spreading1d} shows the time evolution of such correlations in an $L\times L=20\times 20$ system for $|R_x|\leq 8$, with $\alpha=12$. For this $\alpha$, convergence with the number of variational parameters is achieved. Moreover, we expect that correlations at least up to $|R_x|=7$ are free from finite-size effects for the time interval considered here (see Supplemental Material IV). Note also that due to periodic boundary conditions and translation invariance $C(\mathbf{R},t)=C(L\hat{\mathbf{x}}-\mathbf{R},t)$ up to Monte Carlo errors. Fig. \ref{fig:spreading1d} shows that, when considering correlations along one direction, a light-cone like spreading of correlations emerges analogous to what is observed in one-dimensional systems \cite{Calabrese, Verstraete, Alba, Hazzard, Cevolani}. An estimate of the light-cone slope, which gives the spreading speed of correlations, is obtained by fitting the time $t^*$ at which the first extrema appear as a function of $R_x$. In particular, we extract the speed from the inverse of the slope of the fitted line, which characterizes the velocity of correlation propagation between subsequent positions $R_x$. Fig. \ref{fig:maxima}(a) shows the arrival times $t^*$ averaged over positive and negative $R_x$ (red diamonds). The extracted velocity reveals a peculiar bending when going from small to larger distances ($|R_x|>5$), and a spreading speed of the first $|R_x|\leq 5$ correlations of $v(\text{RBM})=(4.71 \pm 0.13)\, J_{\text{ex}}$ (red solid line).

To interpret the results obtained we first turn to linear spin wave theory (LSWT), which is expected to give an accurate account of the correlation spreading for small perturbations and long wavelengths. To this end, the Hamiltonian and perturbation Eqs. (\ref{eq:1}--\ref{eq:2}) are expressed in terms of Holstein-Primakoff boson operators. Next, a Bogolyubov transformation is applied in momentum space such that the resulting linear terms of $\hat{\mathcal{H}}$ are diagonalized. Up to constant terms this yields ($t>0$)
\begin{align}
\hat{\mathcal{H}} + \delta \hat{\mathcal{H}}(t) &= \frac{1}{2} \sum_{\mathbf{k}}  \bigg[\big(\omega_\mathbf{k} + \delta\omega_\mathbf{k}\big)\Big(\hat{\alpha}^\dagger_\mathbf{k}\hat{\alpha}_\mathbf{k} + \hat{\alpha}_{-\mathbf{k}}\hat{\alpha}^\dagger_{-\mathbf{k}}\Big) \nonumber \\ \label{eq:3}
& \hspace{34pt}+ V_\mathbf{k}\Big(\hat{\alpha}^\dagger_\mathbf{k}\hat{\alpha}^\dagger_{-\mathbf{k}}+\hat{\alpha}_\mathbf{k}\hat{\alpha}_{-\mathbf{k}}\Big)\bigg].
\end{align}
Here $\omega_\mathbf{k}$ is the single-magnon dispersion renormalized by the Oguchi factor $Z_c$ \cite{Oguchi}, while $\delta\omega_\mathbf{k}$ and $V_\mathbf{k}$ are proportional to $\Delta J_{\text{ex}}$ and depend on the details of the perturbation. The first term describes the bare magnon spectrum, which is renormalized due to the perturbation of $J_{\text{ex}}$. The second term is responsible for the creation and annihilation of pairs of counter-propagating magnons. In this approximation, the dynamics of spin correlations can be solved analytically, yielding
\begin{equation}\label{eq:lswt_corr}
C(\mathbf{R},t) = C_0(\mathbf{R}) - \frac{1}{N}\sum_{\mathbf{k}} \Gamma_\mathbf{k}\big( e^{i\mathbf{k}\cdot \mathbf{R} +i 2\omega_\mathbf{k}t}+e^{i\mathbf{k}\cdot \mathbf{R} -i 2\omega_\mathbf{k}t}\big),
\end{equation}
in the linear response limit $\Delta J_{\text{ex}}\ll J_{\text{ex}}$,  where $C_0(\mathbf{R})$ is a time-independent term, $\Gamma_{\mathbf{k}}$ is a time-independent factor depending on the geometry of the system and on the perturbation. A detailed derivation is given in Supplemental Material II. 

The spreading speed of the extrema of Eq. \eqref{eq:lswt_corr} is extracted with the same procedure exploited for the RBM correlations. Fig. \ref{fig:maxima}(b) shows the arrival times of the first extrema of Eq. \eqref{eq:lswt_corr} versus the distance $|R_x|$ (black circles). This is compared with the light-cone slope $v_{\text{2M}}$ determined by twice the highest group velocity \cite{Calabrese}, which in the linear response limit is $v_{\text{2M}}=2\frac{d\omega_{\mathbf{k}}}{d\mathbf{k}}|_{\mathbf{k}=0}\approx 3.28 \, J_{\text{ex}}$ (dashed black line). We note that the latter also equals twice the highest phase velocity $\frac{2\,\omega_{\mathbf{k}}}{k_x}\big|_{\mathbf{k}=0}$, which instead determines the spreading speed of the first extrema \cite{Cevolani}. The LSWT results demonstrate that the RBM ansatz yields higher spreading speeds at small times and distances, $v(\text{RBM})$ being more than 40\% higher than the corresponding LSWT speed. Interestingly, the RBM spreading speed at $|R_x|>5$ decreases down to the expected two-magnon velocity as it appears in Fig. \ref{fig:maxima}(a). We refer to the initial regime as \emph{supermagnonic}, since the spreading speed of the correlations is much higher than obtainable from the single-magnon dispersion.

To qualitatively assess the effect of magnon-magnon interactions beyond the Oguchi correction, we consider a Schwinger boson mean-field (SBMFT) representation of the 2D Heisenberg Hamiltonian \cite{Auerbach, Sarker}. This provides an exact solution of the SU($n$) generalization of the Heisenberg Hamiltonian in the limit $n\rightarrow\infty$, in which magnon-magnon interactions remain finite. In addition, opposed to LSWT, SBMFT does not assume a symmetry broken ground state. 

Results for the correlation spreading within SBMFT are shown in Fig. \ref{fig:maxima}(b) (blue squares), where the linear response limit is considered. These results reveal an enhancement of the spreading speed of the first $|R_x|\leq 5$ correlations similar to what is obtained with the RBM ansatz and a speed $v(\text{SBMFT})=(4.06 \pm 0.24)\, J_{\text{ex}}$ (blue solid line). Our calculations show that the SBMFT exhibits two features. Besides propagating modes closely resembling non-interacting two-magnon pairs, an additional quasi-bound state of two spin-flip excitations appears (see Supplemental Material III), similar to what is obtained with other interacting magnon theories \cite{Elliot, Canali, Lorenzana}. The latter dominates the spectrum at small wavelengths, decreasing the frequency of the two-magnon peak as compared to twice the frequency of zone-edge magnons. These two spin-flip excitations are well-known from Raman spectroscopy, since the short-range correlations dominate the Raman spectrum. The supermagnonic propagation, however, is a nontrivial effect arising from the interplay between the quasi-bound state and the propagating magnon pairs. At short distances, the decrease in frequency delays the arrival time of the maxima in Fig. \ref{fig:maxima}(b). This delay rapidly reduces with distance, causing a crossover regime at enhanced speed, recovering the non-interacting propagation speed at large distances.
In a semi-classical picture, this supermagnonic regime can therefore be understood as sub-ballistic propagation of quasi-particles, which only interact when being in close vicinity to each other. As they propagate, the interaction strength decreases and the quasi-particles enter a ballistic regime consistent with the non-interacting two-magnon light cone. We emphasize that SBMFT also encompasses the Oguchi correction that renormalizes the single-magnon spectrum due to magnon-magnon interactions in LSWT. Hence, the appearance of the quasi-bound state results from magnon-magnon interactions between magnon pairs, and goes beyond the Oguchi correction of the single-magnon spectrum.

Within SBMFT, we can further tune the strength of magnon-magnon interactions by varying the spin value $S$. We find that as $S$ increases towards the classical limit $S \rightarrow \infty$, the result converges to that of LSWT (see Supplemental Material III). 
This shows that the significance of the supermagnonic spreading scales with the importance of quantum fluctuations. Overall, the SBMFT results suggest that the RBM data features extraordinary strong magnon-magnon interactions, beyond what can be expected from standard interacting magnon-theories. This analysis is also consistent with the fact that standard interacting magnon theory fails to fully reproduce the exact frequency and width of the spontaneous Raman spectrum of $S=1/2$ antiferromagnets in 2D \cite{Canali, Fabiani, Sandvik 1998}. 

To conclude this section, we comment on the possible experimental verification of the supermagnonic correlation spreading. An interesting material class is the spin-1/2 antiferromagnets comprising copper ions, such as $\text{La}_2\text{CuO}_4$ \cite{Lyons 1998, Coldea, Headings} and CFTD \cite{Christensen, Dalla Piazza}. For example, in CFTD the dominant nearest neighbor exchange is $J_\text{ex}=6.19$ meV. Hence, the fastest two-magnon oscillation period is $T = h/E_\text{max}\approx 160$ fs, where the upper bound for $E_\text{max}=2Z_c\hbar\omega_\text{max}(z-1)/z$ ($z$ being the lattice coordination number) is estimated from interacting magnon theory \cite{Canali}, using the single-magnon energy $\hbar\omega_\text{max}\approx 15$ meV \cite{Christensen}. With a nearest-neighbor distance $a=5.74$ \AA, the supermagnonic velocity is $v\approx 4aJ_{\text{ex}}/\hbar~\sim~20$ km/s, 
and therefore the required experimental resolution is in the nanometer length scale and femtosecond time scale. This is in reach with femtosecond x-ray diffraction techniques, in particular when combined with transient gratings \cite{Bencivenga, Beye, Svetina, Weder}. Moreover, the enhanced spreading speed is also present in systems with higher spin, for example, $S=1$ which enriches the class of materials to fluorides \cite{Zhao, Bossini 2016, Bossini 2019}. 

\section{\emph{Conclusion.}}

In this work we predicted that magnons in two dimensions can propagate with a velocity that is up to 40\% higher than the highest magnon velocity. This supermagnonic speed stems from extraordinary strong magnon-magnon interactions and might be corroborated by femtosecond XFEL experiments with nanometer resolution. Future works might also focus on studying the spreading pattern in detail, both by considering different excitation geometries in the square lattice and by investigating different lattices, such as honeycomb systems. The latter feature even larger quantum fluctuations due to the lower coordination number $z$ \cite{Rosenblum, Kim} and hence may therefore exhibit an enhanced supermagnonic regime. Furthermore, it will be interesting to gain insight in the role of fractionalized spin excitations \cite{Dalla Piazza, Sandvik 2017} and magnon-Higgs scattering \cite{Powalski 2015, Powalski 2018} on the space-time propagation of magnon pairs. Moreover, such propagation is generally accompanied by a linear growth of entanglement \cite{Calabrese}. How this is affected by magnon-magnon interactions is currently under investigation.

This work is part of the Shell-NWO/FOM-initiative ``Computational sciences for energy research” of Shell and Chemical Sciences, Earth and Life Sciences, Physical Sciences, FOM and STW. Part of this work was carried out on the Dutch national e-infrastructure with the support of SURF Cooperative.

\pagebreak
\widetext
\begin{center}
\textbf{\large Supplemental Materials for `Supermagnonic propagation in two-dimensional antiferromagnets'}
\end{center}
\author{G. Fabiani} \author{M. D. Bouman} \author{J. H. Mentink}
\affiliation{Radboud University, Institute for Molecules and Materials (IMM) Heyendaalseweg 135, 6525 AJ Nijmegen, The Netherlands}
\date{\today}

\maketitle
\setcounter{page}{1}
\setcounter{equation}{0}
\setcounter{figure}{0}
\renewcommand{\theequation}{S\arabic{equation}}
\renewcommand\thefigure{S\arabic{figure}}

\section{\textbf{I. Reflection antisymmetry of the correlation function}\\ \\}
In this section, a proof is provided concerning the reflection antisymmetry of the correlation function. We start from the Zubarev formalism \cite{S_Zubarev}, where the time evolution of an observable $\hat{O}$ in response to a time-dependent perturbation $\delta \hat{\mathcal{H}}(t)$ is given by
\begin{equation}
	\langle \hat{O}(t) \rangle = \langle \hat{O} \rangle_0 - \iu
	\int_0^t dt' \big\langle \big[ \hat{O}(t) ,
	\delta \hat{\mathcal{H}}(t') \big] \big\rangle_0 ,
	\label{eq:kubo}
\end{equation}
where the time-dependent operators are taken in the interaction picture, and $\langle . \rangle_0$ denotes the expectation value with respect to the ground state of the unperturbed Hamiltonian $\hat{\mathcal{H}}$. In order to derive the symmetry explicitly, we consider the generic Raman tensor form of the perturbation \cite{S_Fleury,S_Elliot}, which entails an additional Rayleigh scattering term. For the square lattice, this term is $-(\Delta J_{\text{ex}}/4) \sum_{i,\bm{\delta}} \hat{\mathbf{S}}(\mathbf{r}_i) \cdot \hat{\mathbf{S}}(\mathbf{r}_i+\bm{\delta}) \propto \mathcal{H}$, yielding \cite{S_Canali}
\begin{equation}
	\delta\hat{\mathcal{H}}' = \frac{\Delta J_{\text{ex}}}{2} \sum_{i,\bm{\delta}} \bigg[ (\mathbf{e} \cdot \bm{\delta})^2 - \frac{1}{2} \bigg] \hat{\mathbf{S}}(\mathbf{r}_i) \cdot \hat{\mathbf{S}}(\mathbf{r}_i+\bm{\delta}) = \Big(e_x^2-e_y^2\Big) \frac{\Delta J_{\text{ex}}}{2} \sum_i \Big[ \hat{\mathbf{S}}(\mathbf{r}_i) \cdot \hat{\mathbf{S}}(\mathbf{r}_i+\hat{\mathbf{x}}) - \hat{\mathbf{S}}(\mathbf{r}_i) \cdot \hat{\mathbf{S}}(\mathbf{r}_i+\hat{\mathbf{y}}) \Big].
\label{eq:mod_ham}
\end{equation}
Within linear response, this additional term has no effect on the dynamics. This follows from Eq. \eqref{eq:kubo} and the fact that $\big\langle \big[ \hat{O}(t) , \hat{\mathcal{H}}(t') \big] \big\rangle_0 = 0$ for arbitrary $\hat{O}$. The advantage of using $\delta\hat{\mathcal{H}}'$ is that it allows us to factor out the polarization-dependent factor $e_x^2-e_y^2 = \cos 2\phi$, where $\phi$ is the polar angle of the polarization vector $\mathbf{e}$. By Eq. (4) of the main text and Eq. \eqref{eq:kubo}, it follows that the correlation function $C$ is proportional to $\cos 2\phi$. Hence, $C$ factorizes as $C(\mathbf{e},\mathbf{R}) = C_1(\mathbf{e}) C_2(\mathbf{R})$, where $C_1(\mathbf{e})=\cos 2\phi$ and $\mathbf{R}$ the position vector.
We are interested in the behavior of $C$ under the action of a transformation matrix $\mathbf{M}$ on the vectors $\mathbf{e}$ and $\mathbf{R}$, where $\mathbf{M}$ describes a reflection over one of the diagonals of the lattice (or equivalently interchanging $\mathbf{x}$ and $\pm \mathbf{y}$). Evidently we have $C_1(\mathbf{Me}) C_2(\mathbf{MR}) = C_1(\mathbf{e}) C_2(\mathbf{R})$ due to the symmetry of the Hamiltonian and perturbation. In addition, from the specific $\cos 2\phi$-dependence we obtain $C_1(\mathbf{Me}) = - C_1(\mathbf{e})$. Therefore, it follows that $C_2(\mathbf{MR}) = - C_2(\mathbf{R})$. 
Thus, due to the factorization property, the full correlation function possesses diagonal antisymmetry. This is a general consequence of the symmetry of the perturbation, and thus does not depend on the way the spin operators are treated (i.e., exact, SWT, SBMFT, etc.). Beyond linear response, this symmetry is broken, as $C_1(\mathbf{Me}) = - C_1(\mathbf{e})$ no longer holds due to $\mathcal{O}(\Delta J_{\text{ex}}^2)$ terms.

\section{\textbf{II. Correlation dynamics in LSWT}\\ \\}
In this section, the results of the main text concerning linear spin wave theory and the corresponding dynamics of spin correlations are derived. To this end, we perform a unitary transformation on the spin operators, consisting of a $\pi$ rotation about the $\mathbf{y}$-axis of one of the two sublattices defined by the checkerboard decomposition of the square lattice. This yields the transformed operators
\begin{equation} \label{eq:subl_rot}
\hat{\tilde{S}}^z_i = e^{\iu \bm{\pi}\cdot \mathbf{r}_i}\hat{S}^z_i, \hspace{30pt} \hat{\tilde{S}}^x_i = e^{\iu \bm{\pi} \cdot \mathbf{r}_i}\hat{S}^x_i, \hspace{30pt} \hat{\tilde{S}}^y_i = \hat{S}^y_i,
\end{equation}
where $e^{\iu \bm{\pi}\cdot \mathbf{r}_i}=+1\,(-1)$ for sites in the non-rotated (rotated) sublattice. This transformation allows us to define only one species of boson operators. Here we consider the low-order Holstein-Primakoff transformation
\begin{equation} \label{eq:HP}
\hat{\tilde{S}}_{i}^z = S-\hat{a}^\dagger_{i}\hat{a}_{i}, \hspace{30pt} \hat{\tilde{S}}^+_{i} = \sqrt{2S}\hat{a}_{i},\hspace{30pt}  \hat{\tilde{S}}^-_{i} = \sqrt{2S}\hat{a}^\dagger_{i}.
\end{equation}
It is convenient to work in momentum space, where the Holstein-Primakoff bosons are expressed as
\begin{equation}\label{eq:fourier}
\hat{a}_\mathbf{k} = \frac{1}{\sqrt{N}}\sum_i e^{-\iu{\mathbf{k}}\cdot \mathbf{r}_i}\hat{a}_{i}, \hspace{30pt} \hat{a}_i = \frac{1}{\sqrt{N}}\sum_{\mathbf{k}}e^{\iu\mathbf{k} \cdot \mathbf{r}_i}\hat{a}_{\mathbf{k}},
\end{equation}
where the $i$-sum is over the full lattice, and the $\mathbf{k}$-sum is over the full Brillouin zone. By substituting Eqs. (\ref{eq:HP}, \ref{eq:fourier}) into Eq. \eqref{eq:1} of the main text, we obtain
\begin{equation}
\hat{\mathcal{H}} = -\frac{Nz}{2}J_{\text{ex}} S(S+1) + \frac{zJ_{\text{ex}}S}{2}\sum_{\mathbf{k}}\bigg[\hat{a}_{\mathbf{k}}^\dagger\hat{a}_{\mathbf{k}} + \hat{a}_{\mathbf{-k}}\hat{a}^\dagger_{\mathbf{-k}} - \gamma_{\mathbf{k}}\Big(\hat{a}_{\mathbf{k}}^\dagger\hat{a}_{-\mathbf{k}}^\dagger+\hat{a}_{\mathbf{k}}\hat{a}_{-\mathbf{k}}\Big)\bigg]\nonumber,
\end{equation}
where $z$ is the coordination number ($z=4$ for the square lattice), and $\gamma_{\mathbf{k}} = \frac{1}{z}\sum_{\bm{\delta}}e^{\iu\mathbf{k}\cdot \bm{\delta}}$. This Hamiltonian is diagonalized with a Bogoliubov transformation $\hat{\alpha}_{\mathbf{k}} = \cosh \theta_{\mathbf{k}} \,\hat{a}_{\mathbf{k}} - \sinh \theta_{\mathbf{k}} \,\hat{a}^\dagger_{-\mathbf{k}}$, where $\tanh 2\theta_{\mathbf{k}} = \gamma_{\mathbf{k}}$. This yields
\begin{equation}\label{eq:lswt_hamiltonian}
\hat{\mathcal{H}} = -\frac{Nz}{2}J_{\text{ex}} S(S+1) + \frac{1}{2}\sum_{\mathbf{k}}\omega_{\mathbf{k}}\Big(\hat{\alpha}_{\mathbf{k}}^\dagger\hat{\alpha}_{\mathbf{k}} + \hat{\alpha}_{-\mathbf{k}}\hat{\alpha}^\dagger_{-\mathbf{k}} \Big), \hspace{30pt} \omega_{\mathbf{k}}= zSJ_{\text{ex}}\sqrt{1-\gamma^2_{\mathbf{k}}}.
\end{equation}
In the spin wave calculation of the main text we employed the Oguchi correction to the single-magnon spectrum \cite{S_Oguchi}, which is given by the renormalization $\omega_{\mathbf{k}}\rightarrow Z_c \,\omega_{\mathbf{k}}$, where
\begin{equation} \nonumber
    Z_c = 1 + \frac{1}{2S} \frac{1}{N} \sum_{\mathbf{k}} \bigg(1-\sqrt{1-\gamma_{\mathbf{k}}^2}\bigg) \approx 1.158 .
\end{equation}
This captures the most simple effect of magnon-magnon interactions.

We express the perturbation $\delta\hat{\mathcal{H}}(t)$ in terms of the same bosons. This basis is convenient because it allows us to express the initial state at $t=0$ as a vacuum state. However, as a consequence the perturbation is not diagonal in this basis. Up to a constant term, this procedure yields
\begin{equation}
\begin{split}
\delta \hat{\mathcal{H}}(t) &= \frac{1}{2}\sum_{\mathbf{k}}\bigg[\delta\omega_{\mathbf{k}}\Big(\hat{\alpha}_{\mathbf{k}}^\dagger\hat{\alpha}_{\mathbf{k}} + \hat{\alpha}_{-\mathbf{k}}\hat{\alpha}^\dagger_{-\mathbf{k}}\Big)
+ V_{\mathbf{k}}\Big(\hat{\alpha}_{\mathbf{k}}^\dagger\hat{\alpha}_{-\mathbf{k}}^\dagger + \hat{\alpha}_{\mathbf{k}}\hat{\alpha}_{-\mathbf{k}}\Big)\bigg], \\
\delta \omega_{\mathbf{k}} &= zS\Delta J_{\text{ex}}(t) \frac{1-\xi_{\mathbf{k}}\gamma_{\mathbf{k}}}{\sqrt{1-\gamma^2_{\mathbf{k}}}}, \hspace{30pt} V_{\mathbf{k}} = -zS\Delta J_{\text{ex}}(t) \frac{\tau_{\mathbf{k}}\cos 2\phi }{\sqrt{1-\gamma^2_{\mathbf{k}}}} ,
\end{split}
\end{equation}
where we defined $\xi_{\mathbf{k}} = \text{cos}^2 \phi\,\cos k_x +\text{sin}^2\phi \cos k_y$, and the factor $\tau_{\mathbf{k}}=(\cos k_x - \cos k_y)/2$ exhibits an antisymmetry under interchanging $\mathbf{x}$ and $\pm\mathbf{y}$. Next, the expression Eq. \eqref{eq:LSWT_corr} of the main text is derived. For ease of notation we define the two-magnon operators \cite{S_Bossini 2019}
\begin{equation}
\hat{K}^z_{\mathbf{k}} = \frac{1}{2}\Big(\hat{\alpha}_{\mathbf{k}}^\dagger\hat{\alpha}_{\mathbf{k}} + \hat{\alpha}_{-\mathbf{k}}\hat{\alpha}^\dagger_{-\mathbf{k}}\Big), \hspace{30pt}
\hat{K}^+_{\mathbf{k}} = \hat{\alpha}_{\mathbf{k}}^\dagger\hat{\alpha}_{-\mathbf{k}}^\dagger, \hspace{30pt} \hat{K}^-_{\mathbf{k}} = \hat{\alpha}_{\mathbf{k}}\hat{\alpha}_{-\mathbf{k}}. \nonumber
\end{equation}
Subsequently, we consider the definition Eq. \eqref{eq:corr} of the main text and express the correlation function in terms of $\hat{K}_{\mathbf{k}}^z,\hat{K}_{\mathbf{k}}^{\pm}$ in the linear spin wave approximation. This yields
\begin{equation}\label{eq:C_LSWT}
C(\mathbf{R},t)=\frac{2S}{N}\sum_{\mathbf{k}}\frac{\cos(\mathbf{k}\cdot\mathbf{R})}{\sqrt{1-\gamma^2_{\mathbf{k}}}}\bigg[ \big\langle \hat{K}^z_{\mathbf{k}}\big\rangle + \frac{\gamma_{\mathbf{k}}}{2}\Big( \big\langle \hat{K}^+_{\mathbf{k}}\big\rangle+ \big\langle \hat{K}^-_{\mathbf{k}}\big\rangle\Big)-\frac{1}{2}\bigg],
\end{equation}
for $\mathbf{R}$ connecting spins in the same sublattice; a similar expression can be obtained when $\mathbf{R}$ connects different sublattice spins. The dynamics of the correlation function is determined solely by the time evolution of $\big\langle\hat{K}_{\mathbf{k}}^z\big\rangle$, $\big\langle\hat{K}^{\pm}_{\mathbf{k}}\big\rangle$. This can be obtained by solving the Heisenberg equations of motion for the expectation values
\begin{equation}
\frac{d\big\langle\hat{K}^z_{\mathbf{k}}\big\rangle}{dt}= \iu V_{\mathbf{k}}\Big[\big\langle\hat{K}^-_{\mathbf{k}}\big\rangle-\big\langle\hat{K}^+_{\mathbf{k}}\big\rangle\Big], \hspace{30pt}
\frac{d\big\langle\hat{K}^{\pm}_{\mathbf{k}}\big\rangle}{dt}= \pm 2 \iu\Big[(\omega_{\mathbf{k}}+\delta\omega_{\mathbf{k}})\big\langle\hat{K}^{\pm}_{\mathbf{k}}\big\rangle + V_{\mathbf{k}}\big\langle\hat{K}^z_{\mathbf{k}}\big\rangle\Big], \nonumber
\end{equation}
with initial conditions given by the ground state values $\big\langle\hat{K}^z_{\mathbf{k}}\big\rangle = \frac{1}{2}$, $\big\langle\hat{K}^\pm_{\mathbf{k}}\big\rangle = 0$. For the perturbation $\Delta J_{\text{ex}}(t)=0.1\, J_{\text{ex}}\,\Theta(t)$ as considered in the main text, these equations can be solved analytically, yielding (for $t>0$)
\begin{equation}
\big\langle\hat{K}_{\mathbf{k}}^z\big\rangle = \frac{V_{\mathbf{k}}^2}{2 b_{\mathbf{k}}^2}\Big[1-\cos(2 b_{\mathbf{k}}t)\Big]+\frac{1}{2}, \hspace{30pt}
\big\langle\hat{K}_{\mathbf{k}}^+\big\rangle + \big\langle\hat{K}_{\mathbf{k}}^-\big\rangle = -\frac{V_{\mathbf{k}}(\omega_{\mathbf{k}}+\delta\omega_{\mathbf{k}})}{b_{\mathbf{k}}^2}\Big[1-\cos(2 b_{\mathbf{k}}t)\Big],\nonumber
\end{equation}
where $b_{\mathbf{k}} = \sqrt{(\omega_{\mathbf{k}}+\delta\omega_{\mathbf{k}})^2-V_{\mathbf{k}}^2}$ is the single-magnon spectrum of the quenched Hamiltonian, i.e., the time-independent Hamiltonian $\hat{\mathcal{H}}+\delta\hat{\mathcal{H}}(t)$ at $t>0$ (not explicitly derived here). Inserting these solutions into Eq. \eqref{eq:C_LSWT} yields
\begin{equation}\label{eq:LSWT_corr}
\begin{split}
&\hspace{20pt}C(\mathbf{R},t)=C_0(\mathbf{R})-\frac{1}{N}\sum_{\mathbf{k}}\Gamma_{\mathbf{k}}\Big(e^{\iu\mathbf{k}\cdot\mathbf{R}+\iu2 b_{\mathbf{k}}t}+e^{\iu\mathbf{k}\cdot\mathbf{R}-\iu 2 b_{\mathbf{k}}t}\Big), \\
\text{with}\hspace{20pt} &C_0(\mathbf{R}) = \frac{1}{N}\sum_{\mathbf{k}}\Gamma_{\mathbf{k}}\cos(\mathbf{k}\cdot \mathbf{R}),\hspace{30pt}   
\Gamma_{\mathbf{k}}=S e^{-2\theta_{\mathbf{k}}} \frac{V_{\mathbf{k}}}{2 b^2_{\mathbf{k}}} \big(\omega_{\mathbf{k}}+\delta\omega_{\mathbf{k}}+V_{\mathbf{k}}\big).
\end{split}
\end{equation}

In the linear response approximation, only terms of the order $\Delta J_{\text{ex}}$ are kept, which yields $\Gamma_{\mathbf{k}}=S\exp(-2\theta_{\mathbf{k}}) V_{\mathbf{k}}/2\omega_{\mathbf{k}}$. Thus in this case, $\Gamma_{\mathbf{k}}$ inherits its symmetry from $V_{\mathbf{k}}$, which in turn stems from $\tau_{\mathbf{k}}$. Specifically, this entails antisymmetry under reflections over one of the two diagonals of the lattice. As a consequence, $C(\mathbf{R},t)$ also possesses this antisymmetry, which is in agreement with the findings of section I. This antisymmetry has a crucial effect on the correlation spreading, as $C({\mathbf{R}},t)$ vanishes on the diagonals in the linear response limit. Beyond linear response, correlations are allowed to spread along the diagonals. However, the amplitude is of the order $\Delta J_{\text{ex}}^2$ and therefore small for the value $\Delta J_{\text{ex}}=0.1 J_{\text{ex}} $ of the main text.

\section{\textbf{III. Correlation dynamics in SBMFT}\\ \\}
In this section, the results of the main text concerning Schwinger boson mean-field theory and the corresponding dynamics of spin correlations are derived. Firstly, we treat the standard static SBMFT. Secondly, we show how SBMFT can be used to calculate dynamics of observables in the linear response limit. Finally, we apply this to the spin correlation function and analyze the result in both the spectral representation and in real time.
\subsection{A. SBMFT}
In the Schwinger boson representation, the spin operators are described by bosons with two `flavors' indicated by the label $s=\pm1/2$:
\begin{equation} \nonumber
	\hat{\tilde{S}}_i^z = \frac{1}{2} \Big( \hat{a}^{\dag}_{i,\frac{1}{2}}
	\hat{a}_{i,\frac{1}{2}} - \hat{a}^{\dag}_{i,-\frac{1}{2}}
	\hat{a}_{i,-\frac{1}{2}} \Big) ,
	\hspace{30pt}
	\hat{\tilde{S}}_i^- = \hat{a}_{i,-\frac{1}{2}}^{\dag} \hat{a}_{i,\frac{1}{2}} ,
	\hspace{30pt}
	\hat{\tilde{S}}_i^+ = \hat{a}^{\dag}_{i,\frac{1}{2}} \hat{a}_{i,-\frac{1}{2}} ,
\end{equation}
where the sublattice rotation of Eq. \eqref{eq:subl_rot} is employed. Whereas the HP bosons are described by a single flavor in a fixed spin-subspace, the two Schwinger boson flavors can create states in any spin-subspace. To select a single subspace with given $S$, the total number of bosons on a site is constrained according to $\sum_s \hat{a}^{\dag}_{i,s} \hat{a}_{i,s} = 2S$. This is enforced by a Lagrange multiplier. Up to a constant term, the static mean-field Hamiltonian in momentum space reads \cite{S_Auerbach,S_Sarker}
\begin{equation} \nonumber
	\hat{\mathcal{H}} = \frac{1}{2} \sum_{\mathbf{k}, s}
	\bigg[ \lambda_0 \Big( \hat{a}^{\dag}_{\mathbf{k},s} \hat{a}_{\mathbf{k},s}
	+ \hat{a}_{-\mathbf{k},s} \hat{a}^{\dag}_{-\mathbf{k},s} \Big)
	- z Q_0 \gamma_{\mathbf{k}} \Big( \hat{a}^{\dag}_{\mathbf{k},s}
	\hat{a}^{\dag}_{-\mathbf{k},s} + \hat{a}_{-\mathbf{k},s}
	\hat{a}_{\mathbf{k},s} \Big) \bigg] ,
\end{equation}
where the momentum-space operators are defined analogous to Eq. \eqref{eq:fourier}. There are two mean-field parameters: the Lagrange multiplier $\lambda_0$, and the bond parameter $Q_0$ defined in the right equation of Eq. (\ref{eq:SC}), where $i,j$ are nearest neighbors. Both of these are taken to be uniform throughout the lattice. Due to a U(1) gauge symmetry, the bond parameter can be chosen to be real. $\hat{\mathcal{H}}$ is diagonalized with a Bogoliubov transformation $\hat{\alpha}_{\mathbf{k},s} = \cosh \theta_{\mathbf{k}} \,\hat{a}_{\mathbf{k},s} - \sinh \theta_{\mathbf{k}} \,\hat{a}^{\dag}_{-\mathbf{k},s}$, where $\tanh 2\theta_{\mathbf{k}} = zQ_0\gamma_{\mathbf{k}}/\lambda_0$, yielding
\begin{equation}
	\hat{\mathcal{H}} = \frac{1}{2} \sum_{\mathbf{k},s}
	\omega_{\mathbf{k}} \Big( \hat{\alpha}_{\mathbf{k},s}^{\dag} \hat{\alpha}_{\mathbf{k},s}
	+\hat{\alpha}_{-\mathbf{k},s} \hat{\alpha}_{-\mathbf{k},s}^{\dag} \Big),
	\hspace{30pt}
	\omega_{\mathbf{k}} = \sqrt{\lambda_0^2 - (zQ_0\gamma_{\mathbf{k}})^2} .
\end{equation}
The mean-field parameters are determined by a set of self-consistent equations. These are obtained by enforcing the expectation value of the constraint on each site, and by expressing the bond parameter self-consistently:
\begin{equation}
	2S = \sum_s \big\langle \hat{a}^{\dag}_{i,s} \hat{a}_{i,s} \big\rangle ,
	\hspace{30pt}
	Q_0 = \frac{J_{\mathrm{ex}}}{2} \sum_s \big\langle
	\hat{a}_{i,s} \hat{a}_{j,s} \big\rangle .
\label{eq:SC}
\end{equation}
Evaluating these expectation values on the ground state of $\hat{\mathcal{H}}$ yields the self-consistent equations. The resulting dispersion $\omega_{\mathbf{k}}$ has very close agreement with the LSWT result Eq. \eqref{eq:lswt_hamiltonian}, provided the Oguchi correction is included in LSWT. This indicates that SBMFT captures the same single-magnon frequency renormalization as the Oguchi correction. Note that we only consider finite systems, such that there is no spontaneous symmetry breaking by Bose condensation.

\subsection{B. Self-consistent dynamics in linear response}
Next we discuss a general framework to calculate linear response dynamics in SBMFT self-consistently (see \cite{S_Bouman thesis, S_Bouman} for a detailed derivation). Compared to LSWT, calculating dynamics in SBMFT is more involved, since the definition of the mean-field parameters changes in the presence of a perturbation. In order to obtain a self-consistent result at all instances of time, a set of dynamical self-consistent equations is constructed that determines the time-dependence of the mean-field parameters. This is done up to first order in the perturbation strength, for which we define the smallness parameter as $\varepsilon = \frac{2}{z} \Delta J_{\mathrm{ex}} / J_{\mathrm{ex}}$. Contrary to the static solution, two different bond parameters $Q_x$ and $Q_y$ are needed since the perturbation breaks the four-fold symmetry of the lattice. The self-consistent equations are then obtained analogous to Eq. \eqref{eq:SC}, and using Eq. \eqref{eq:kubo} to evaluate the expectation value dynamically. For the step-like perturbation of the main text, the resulting set of equations can be solved analytically by applying a Laplace transformation, yielding (for $t>0$)
\begin{equation} \nonumber
	\lambda(t) = \lambda_0 ( 1 + \varepsilon ) ,
	\hspace{30pt}
	Q(t) = Q_0 ,
	\hspace{30pt}
	q(t) = \varepsilon Q_0 \big[ q_{\mathrm{r}}(t)
	+ \iu q_{\mathrm{i}}(t) \big] \cos(2\phi) .
\end{equation}
Here we defined $Q = (Q_x + Q_y)/2$, $q = (Q_x - Q_y)/2$, which separates the terms that are symmetric and antisymmetric with respect to diagonal reflections. The functions $q_{\mathrm{r}}(t)$ and $q_{\mathrm{i}}(t)$ are known analytically in the Laplace domain and comprise damped oscillations. The perturbation can be written as
\begin{equation} \label{eq:SBMFT_perturbation}
	\delta \hat{\mathcal{H}}(t) = \frac{1}{2} \sum_{\mathbf{k},s} \Big[ V_{\mathbf{k}}(t)
	\hat{\alpha}_{\mathbf{k},s}^{\dag} \hat{\alpha}_{-\mathbf{k},s}^{\dag}
	+ V_{\mathbf{k}}^*(t) \hat{\alpha}_{\mathbf{k},s}
	\hat{\alpha}_{-\mathbf{k},s} \Big] ,
	\hspace{25pt}
	V_{\mathbf{k}}(t) =
	- \varepsilon z Q_0 \tau_{\mathbf{k}} \cos(2\phi) 
	\bigg\{ \frac{\lambda_0}{\omega_{\mathbf{k}}} \big[1 + q_{\mathrm{r}}(t)\big]
	+ \iu q_{\mathrm{i}}(t) \bigg\} .
\end{equation}
These expressions can be used to calculate the dynamics of an observable analytically in the Laplace domain. The time-domain solution is finally obtained by evaluating the inverse Laplace transform numerically. 

\subsection{C. Correlation dynamics}
To study the dynamics of the correlation function, firstly an expression for the correlation function is derived, and secondly the result is studied in momentum and frequency space and subsequently in real space and time. Applying Eqs. (\ref{eq:kubo}, \ref{eq:SBMFT_perturbation}) to Eq. \eqref{eq:corr} of the main text yields
\begin{equation} \label{eq:C_sbmft}
\begin{split}
	& \hspace{40pt} C(\mathbf{R},t) = \varepsilon \bigg[ \frac{1}{N}
	\sum_{\mathbf{k}} e^{\iu \mathbf{k} \cdot \mathbf{R}}
	e^{-2 \theta_{\mathbf{k}}} \bigg] \bigg[ \frac{1}{N}
	\sum_{\mathbf{k}} e^{\iu \mathbf{k} \cdot \mathbf{R}}
	e^{2 \theta_{\mathbf{k}}} W_{\mathbf{k}}(t) \bigg] ,
	\\
	W_{\mathbf{k}}(t) &= \frac{z Q_0 \lambda_0}{\omega_{\mathbf{k}}}
	\tau_{\mathbf{k}} \cos(2\phi) \int_0^t dt' \,
	p(t') \, \mathrm{sin}\big[ 2 \omega_{\mathbf{k}} (t-t') \big] ,
	\hspace{30pt}
	p(t) = 1+ q_{\mathrm{r}}(t) - \frac{1}{2\lambda_0}
	\frac{ d q_{\mathrm{i}}}{dt} .
\end{split}
\end{equation}
The functional form of Eq. \eqref{eq:C_sbmft} allows for a natural separation of the correlation function into two terms: $C_{\text{stat}} \, (W_{\mathbf{k},\text{stat}})$ corresponding to the static part of $p(t)$, and $C_{\text{dyn}} \, (W_{\mathbf{k},\text{dyn}})$ corresponding to its dynamical part. Here, the static part refers to the value in the limit $t \rightarrow \infty$, when the transient oscillations have damped. The analytical expression for the Laplace transform of $W_{\mathbf{k}}(t)$ reads
\begin{figure}
    \vspace*{-5pt}\centering
    \subfloat{{\includegraphics[scale=0.8]{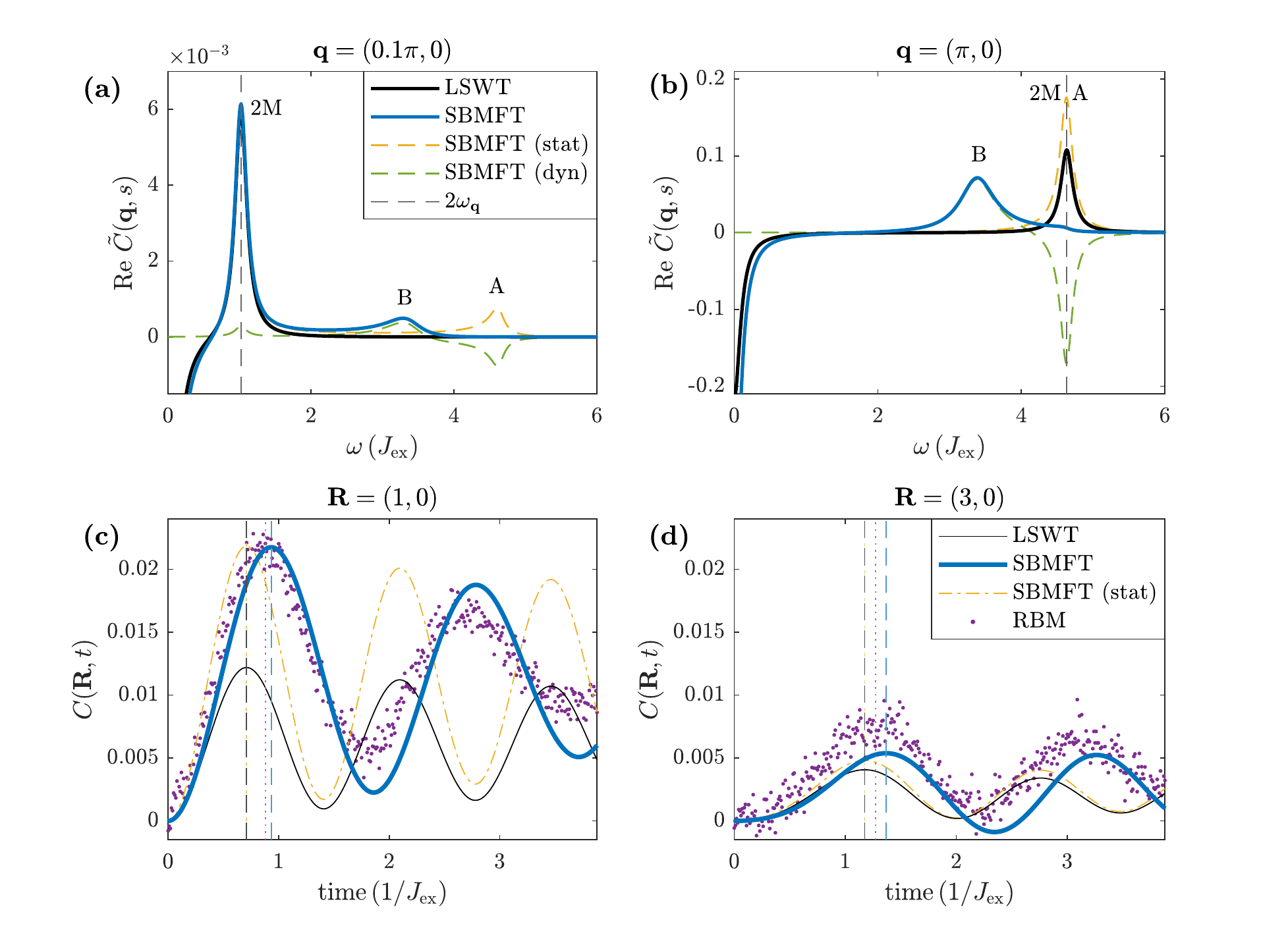}}}
    \caption{(a)--(b) Spectral representation of the correlation function as obtained from the Laplace transform by setting $s = 0.1\,J_{\text{ex}}+\iu \omega$ in an $L \times L = 140 \times 140$ system. The solid black (blue) line represents the LSWT (SBMFT) result. The dashed orange (green) line represents SBMFT with the static (dynamical) part of the mean-field parameters. The vertical black dashed line indicates twice the single-magnon frequency, which is practically identical for LSWT and SBMFT.
    (c)--(d) Real space and time dynamics of the correlation function in an $L \times L = 20 \times 20$ system. The black, blue, and yellow lines represent LSWT, SBMFT, and SBMFT with static mean-field parameters, respectively. The purple dots represent RBM. The vertical lines indicate the positions of the first extrema, with color corresponding to the different methods.
    All results are obtained within linear response theory. The SBMFT (stat) result agrees closely with the LSWT result. Including SBMFT (dyn) yields an additional peak at $\omega = 3.3$--$3.4\, J_{\text{ex}}$, which is dominant at high $\mathbf{q}$. In the real space and time dynamics, this new peak drastically alter the dynamics at short distances, reducing the oscillation frequency and giving closer agreement with RBM.}
    \label{fig:sbmft}
\end{figure}
\begin{equation} \label{eq:SBMFT_W}
\begin{split}
	& \hspace{20pt} \tilde{W}_{\mathbf{k}}(s) = 2 z Q_0 \lambda_0 \cos(2\phi)
	\frac{\tau_{\mathbf{k}}}{(2\omega_{\mathbf{k}})^2+s^2}
	\frac{1}{s} \frac{1}{1-\tilde{K}(s)	\big[1-(s/\bar{s} )^2\big]} ,
	\\
	\tilde{K}(s) &= \frac{J_{\mathrm{ex}}}{N} \sum_{\mathbf{k}}
	\frac{z \lambda_0^2 \tau_{\mathbf{k}}^2}
	{\omega_{\mathbf{k}}^2} \frac{2\omega_{\mathbf{k}}}
	{(2\omega_{\mathbf{k}})^2+s^2} ,
	\hspace{30pt}
	\bar{s} = 2\lambda_0 - \frac{J_{\mathrm{ex}}}{N} \sum_{\mathbf{k}}
	\frac{z \lambda_0\tau_{\mathbf{k}}^2}{\omega_{\mathbf{k}}}
	= 3.95 \, J_{\mathrm{ex}} .
\end{split}
\end{equation}
Similar to LSWT and the exact linear response result, the correlation function has diagonal antisymmetry due to the factor $\tau_{\mathbf{k}}$.

To study the spectral representation of the correlation function, we perform a spatial Fourier transform and evaluate $\tilde{C}(\mathbf{q},s)$ at $s = 0.1\,J_{\text{ex}}+\iu \omega$ in the Laplace domain. The result is shown in Figs. \ref{fig:sbmft}(a)--(b), and is compared with LSWT in linear response (see Eq. \eqref{eq:LSWT_corr}). We observe that LSWT and SBMFT (stat) agree closely as both exhibit a main two-magnon peak at twice the single-magnon frequency $2 \omega_{\mathbf{q}}$ (indicated by `2M'), describing the excitation of a pair of counter-propagating non-interacting magnons. In addition to the 2M peak, SBMFT (stat) exhibits a peak at $\omega = 2\lambda_0$ (indicated by `A'). By including the dynamical contribution, this peak is shifted to a lower frequency $\omega= $ 3.3--3.4$\, J_{\text{ex}}$ (indicated by `B'), and is broadened to a width of approximately $0.2 \, J_{\text{ex}}$. For large $\mathbf{q}$, the dynamical contribution also suppresses the 2M peak, and peak B dominates the spectrum. The excitations corresponding to `A' and `B' are always associated with the zone-edge magnons, which indicates a localized character. We ascribe both peaks to a localized double spin-flip excitation, distinct from the propagating two-magnon excitations. Our interpretation is that the dynamical contribution captures the magnon-magnon interactions beyond the Oguchi correction, which allow for the formation of a quasi-bound state of such a double excitation. Thereby, both its energy and life time are reduced. These observations agree with what is known from spontaneous Raman (SR) spectroscopy \cite{S_Elliot, S_Canali, S_Lorenzana}, where it is well-established that magnon-magnon interactions lead to a two-magnon quasi-bound state with lower energy. Following the argument by \cite{S_Canali,S_Devereaux}, this quasi-bound state can be intuitively understood as follows. The perturbation flips the spins in neighboring sites. In the Ising limit this process costs an energy $3 \, J_{\text{ex}}$, as six nearest-neighbor bonds are broken. By taking into account the Oguchi correction, we arrive at an excitation energy $3 Z_c J_{\text{ex}} \approx 3.4 \, J_{\text{ex}}$, which agrees with the SBMFT result.

\begin{figure}[t]
    \vspace*{-15pt}\centering
    \subfloat{{\includegraphics[scale=0.85]{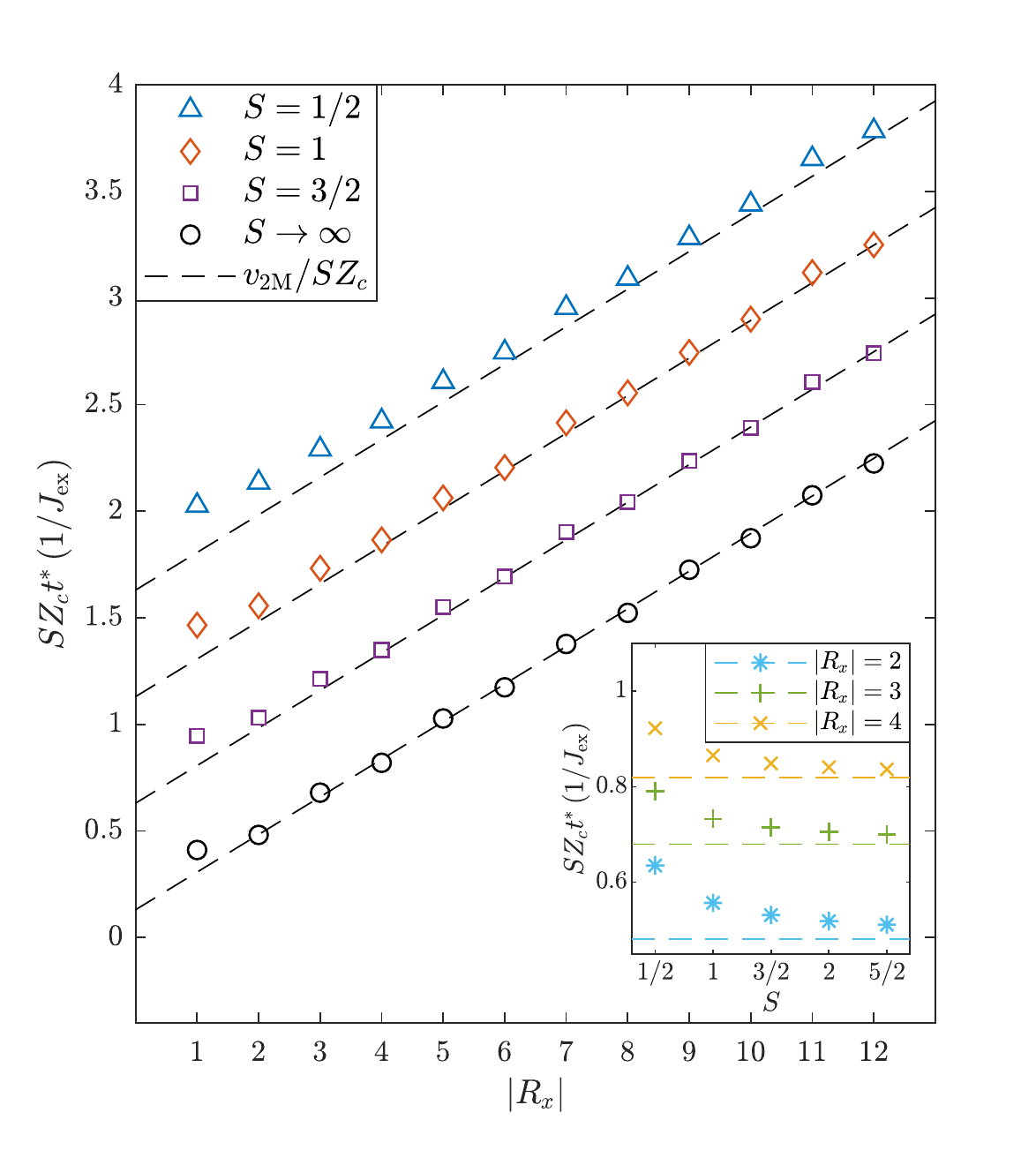}}}
    \caption{Arrival times $t^*$ of the first extrema of $C(\mathbf{R},t)$ in SBMFT as a function of distance $R_x$, for spin values $S=1/2$ (blue triangles), $S=1$ (red diamonds), $S=3/2$ (purple squares), and the limit $S \rightarrow \infty$ (black circles) which coincides with LSWT. Data are compared with the highest magnon group velocity $v_{\text{2M}}$ (black dashed line). To improve readability, the $S=3/2$, $S=1$ and $S=1/2$ data have been shifted in time by $+0.5/J_{\text{ex}}$, $+1/J_{\text{ex}}$, and $+1.5/J_{\text{ex}}$ respectively. Inset: Arrival times $t^*$ as a function of spin $S$, for distance values $|R_x| = 2$ (light-blue asterisks), $|R_x| = 3$ (green pluses), and $|R_x| = 4$ (yellow crosses). The colored dashed lines refer to the large-$S$ limit given by LSWT, with color corresponding to $|R_x|$. All data refer to an $L\times L=52\times 52$ system, where finite-size convergence is found up to $|R_x|=12$. In these figures, $t^*$ is renormalized by a factor $SZ_c$ such that the result in the non-interacting LSWT limit is independent of spin. As $S$ increases, the result converges to the large-$S$ limit given by LSWT.}
    \label{fig:sbmft_S}
\end{figure}

Next we study the real-time dynamics, which is obtained by a numerical inverse Laplace transform of Eq. \eqref{eq:SBMFT_W}. The spatial pattern of spin correlations is qualitatively very similar to LSWT in Fig. \ref{fig:spreading} of the main text (SBMFT data not shown). Figs. \ref{fig:sbmft}(c)--(d) show the temporal profile of the correlation function, and a comparison with LSWT in linear response and RBM. The positions of the first extrema are indicated with vertical lines. The LSWT and SBMFT (stat) results agree very closely, except for a difference in amplitude. In the case of full SBMFT, the oscillation frequency is drastically reduced for small $\mathbf{R}$, and closer agreement with RBM is obtained. At large $\mathbf{R}$, all curves coincide more closely (data not shown). These observations are a direct result of the quasi-bound state peak dominating at high (zone-edge) momenta, which correspond to small distances. This lower oscillation frequency causes a delay in the arrival time of the first extrema at small $\mathbf{R}$, and this delay is reduced as $\mathbf{R}$ increases. This explains the bending observed in Fig. \ref{fig:maxima} of the main text. Thus for SBMFT we also observe a supermagnonic propagation velocity. In this case, the origin of the supermagnonic effect can be directly traced back to the existence of the quasi-bound state, caused by the magnon-magnon interactions on the two-particle level and hence go beyond the Oguchi correction which only renormalizes the single magnon spectrum. 

We note that there are still discrepancies between the RBM and SBMFT results, as in the latter the spreading speed is lower and the supermagnonic regime extends to longer-distance correlations. We exclude the presence of numerical errors in the RBM results: for $\alpha=12$, convergence with the number of variational parameters is achieved; moreover, we expect that correlations at least up to $|R_x|=7$ are free from finite-size effects for the time interval considered here (see Supplemental Material IV). Therefore we ascribe these discrepancies to the intrinsic errors in applying a large-$n$ theory to the SU($n=2$) Heisenberg Hamiltonian. Specifically, they may be related to an underestimation of the width of the quasi-bound state peak, or limited accuracy of the high-energy magnon spectrum. Both of these are not accurately reproduced in interacting magnon theories based on the random phase approximation, as is known from SR spectroscopy \cite{S_Sandvik,S_Wang}. Improvements are also expected by going beyond the linear response approximation employed in the SBMFT calculations.

Finally, we study the spin-dependence of the supermagnonic effect within SBMFT. In the spectral representation (data not shown), the quasi-bound state peak increases in frequency and decreases in width as $S$ increases, in agreement with \cite{S_Elliot}. Fig. \ref{fig:sbmft_S} shows the dependence of the velocity fitted over $|R_x|<5$ for several values of spin $S$. In LSWT, the single-magnon frequencies scale with $SZ_c$ (see Eq. \eqref{eq:lswt_hamiltonian}). Therefore, the arrival times $t^*$ are renormalized by this factor such that the LSWT result is independent of spin. We observe that the supermagnonic effect is most pronounced for $S=1/2$, which features the strongest quantum fluctuations. For large $S$, the amplitude and spatial width of the supermagnonic regime decreases, and the result rapidly converges to non-interacting LSWT. This is a consequence of the fact that the dynamical contribution has a negligible effect in the large-$S$ limit. The observed $S$-dependence illustrates that the strength of magnon-magnon interactions determines the significance of the supermagnonic regime. In the RBM data the supermagnonic propagation is even faster. In addition, for two-magnon excitations the effect of magnon-magnon interactions is already strong as compared to their effect on single-magnon excitations. Therefore, we interpret that the enhanced supermagnonic velocity in the RBM results originates from exceptionally strong magnon-magnon interactions.\\ \\ 

\begin{figure}[t]
    \vspace*{1pt}\centering
    \subfloat{{\includegraphics[width=14.0cm]{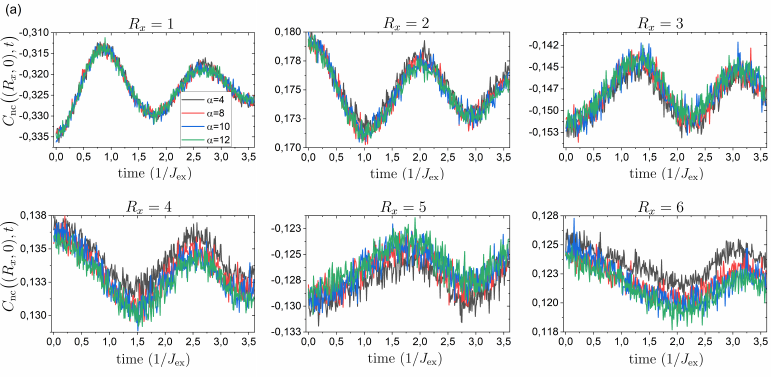} }}\\
    \subfloat{{\includegraphics[width=14.0cm]{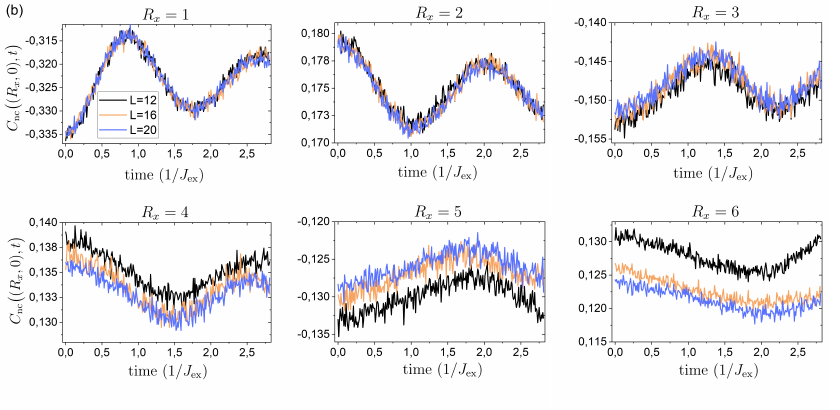} }}\\
    \caption{(a) Time evolution of spin correlations $C_{\text{nc}}(\mathbf{R},t)$ as a function of distance $R_x$ in a $20\times 20$ system for $\alpha=4$ (black line), $\alpha=8$ (red line), $\alpha=10$ (blue line), $\alpha=12$ (green line) and $R_x=$ 1--6. (b) Finite-size scaling of spin correlations $C_{\text{nc}}(\mathbf{R},t)$ for $R_x=$ 1--6. Data are shown for $L=12$ (black line), $L=16$ (orange line) and $L=20$ (blue line). Noise is due to Monte Carlo errors in the evaluation of $C_{\text{nc}}(\mathbf{R},t)$. For the Monte Carlo estimation of the correlations $5\times 10^5$ states have been used. }
    \label{fig:figNQS}
\end{figure}

\section{\textbf{IV. Convergence of the RBM numerical data}\\ \\}
In this section, we show both the dependence on the number of variational parameters, regulated by $\alpha$, and the finite-size scaling of the RBM results of the main text. In particular, we focus on the time evolution of $C_{\text{nc}}(\mathbf{R},t)\equiv \big\langle \hat{\mathbf{S}}(\mathbf{r}_i)\cdot \hat{\mathbf{S}}(\mathbf{r}_i+\mathbf{R})\big\rangle$ and we show that this quantity is numerically converged in an $L\times L=20\times 20$ system at $\alpha=12$ and for the correlations considered in the main text. 

The $\alpha$-dependence of $C_{\text{nc}}(\mathbf{R},t)$ with $\mathbf{R}=(R_x,0)$ is shown in Fig. \ref{fig:figNQS}(a) for an $L\times L=20\times 20$ system and $R_x\in [1,6]$. For such a system, convergence is reached already at $\alpha=10$, as the numerical evaluation of $C_{\text{nc}}(\mathbf{R},t)$ does not change by increasing $\alpha$ within the Monte Carlo noise. Note that large $\mathbf{R}$ correlations are harder to simulate and this is revealed by the slower convergence with $\alpha$.

Fig. \ref{fig:figNQS}(b) shows the finite-size scaling of $C_{\text{nc}}(\mathbf{R},t)$ for systems with $L=12,\, 16, \, 20\, $ and respectively $\alpha=16,\, 16,\, 12$. For all three system sizes the $\alpha$-converged result is considered. We note that convergence of the first extremum of $C_{\text{nc}}(\mathbf{R},t)$ is guaranteed only up to $|R_x|=5$, as for larger distances boundary effects influence the dynamics of the $L=12$ and $L=16$ systems. To which extent the finite size affects the dynamics of correlations at $|R_{x}|>5$ in the $L\times L=20\times 20$ system is not known as this would require the knowledge of the dynamics of larger systems. However, from the scaling shown in Fig. \ref{fig:figNQS}(b), it can be expected that for such a system correlations have converged at least up to $|R_{x}|=7$.

\typeout{get arXiv to do 4 passes: Label(s) may have changed. Rerun}
\end{document}